\documentclass[english]{revtex4-1}
\usepackage[T1]{fontenc}
\usepackage[latin9]{inputenc}
\usepackage{amsmath}
\usepackage{amssymb}

\makeatletter
\usepackage{babel}

\makeatother

\usepackage{babel}
\begin{document}

\title{Finite second-order Born term for Coulomb wavepacket scattering}

\author{Scott E. Hoffmann}

\address{School of Mathematics and Physics,~~\\
 The University of Queensland,~~\\
 Brisbane, QLD 4072~~\\
 Australia}
\email{scott.hoffmann@uqconnect.edu.au}

\selectlanguage{english}%
\begin{abstract}
It has been known for some time that, for nonrelativistic Coulomb
scattering, the terms in the Born series of second and higher order
diverge when using the standard method of calculation. In this paper
we take the matrix elements between square-integrable wavepacket state
vectors. We reproduce the Rutherford cross section from the first-order
contribution. We find that the second-order contribution is finite
and negligible compared to the first-order contribution, away from
the forward direction. At first order, the contribution to the amplitude
in the forward direction is found to be finite and physically reasonable.
We comment on how a similar procedure applied to the divergences of
quantum field theories might render them finite.
\end{abstract}
\maketitle

\section{Introduction}

The Born approximation (or time-dependent perturbation theory) is
the most widely used method for calculating scattering cross sections
in nonrelativistic quantum mechanics \cite{Messiah1961,Cohen-Tannoudji1977,Sakurai1994}
and, with modifications, in relativistic quantum field theories \cite{Itzykson1980}.
In nonrelativistic quantum mechanics it gives an approximation of
the full evolution operator as a series in powers of the potential.
It is only useful, meaning that few terms are needed for a result
of sufficient precision, when the interaction is weak, but this includes
a wide variety of cases.

It has been known for a long time, as illustrated by the calculation
of Dalitz \cite{Dalitz1951}, that for nonrelativistic Coulomb scattering
the terms in the Born series of second and higher order diverge when
using the standard method of calculation. As is well known, the same
pattern of finiteness and divergences happens in relativistic quantum
field theories. We will comment later on the possibility that the
results obtained here can be extended to that domain. Before an investigation
into the divergences of quantum electrodynamics, which will be the
subject of a future work, we consider nonrelativistic Coulomb scattering.
For that system, exact results \cite{Messiah1961,Hoffmann2017a} are
available for comparison. The existence of exact solutions gives us
information about the possibility of a power series solution. The
exact Coulomb scattering amplitude \cite{Messiah1961} depends on
$\Gamma(1+i\eta),$ where $\eta$ is defined below in Eq. (\ref{eq:22}).
Because $\Gamma(1+i\eta)$ has a pole nearest to to the real axis
at $\eta=i,$ we know that there is a series for this amplitude in
powers of $\eta$ that converges only on $|\eta|<1.$ It is this series
that we seek.

It is the aim of this paper to show that the second-order term (for
nonrelativistic Coulomb scattering) becomes finite when square-integrable
wavepackets are used instead of plane waves for the initial and final
states, when the interaction time remains finite rather than infinite
and when care is paid to the order in which integrals are performed.

The results of Dalitz \cite{Dalitz1951} are in terms of an arbitrary,
nonzero, regularization parameter with no physical meaning. The results
diverge as this parameter is taken to zero. Our results will be finite
partly because they depend on a nonzero parameter with physical meaning,
the width in momentum space, $\sigma_{p},$ of the wavepackets. For
simplicity, we choose wavepackets with a spherically symmetric distribution
of momentum about the mean value. This width parameter is not the
eigenvalue of an Hermitian observable, but it is measurable in the
following sense. A state would have to be prepared the same way many
times and momentum measurements performed. The distribution of these
measurements allows calculation of the standard deviation in any direction.

Our construction of an ideal theoretical scattering experiment differs
significantly from the standard method \cite{Newton1982}, in that
a limit as interaction time goes to infinity is replaced by a sequence
where the momentum width is made ever smaller, but never vanishes.
This latter method was discussed in a previous work by this author
\cite{Hoffmann2017a}. We define a small dimensionless parameter
\begin{equation}
\epsilon=\frac{\sigma_{p}}{p},\label{eq:1}
\end{equation}
where $p$ is the magnitude of the initial and final momenta. (We
acknowledge that unperturbed energy is conserved in a scattering description,
so we do not feel the need to prove this fact by choosing different
energies for our initial and final states.) It is important to ensure
that wavepacket spreading is negligible over the finite time of the
interaction. A convenient choice to ensure this is to make the initial
and final separations of the position wavepackets equal to
\begin{equation}
2R=\frac{2}{\sqrt{\epsilon}}\,\sigma_{x},\label{eq:2}
\end{equation}
where $\sigma_{x}\sigma_{p}=1/2$ for minimal Gaussian wavepackets
(setting $\hbar=1$). (We choose Gaussian wavepackets to make the
subsequent integrals tractable. It is beyond the scope of this paper
to investigate the effects of wavepacket shapes on our results.)

Then as $\epsilon$ is made smaller, the resolutions of the momenta
improve, the wavepacket spatial widths increase, but their initial
and final separations increase at a greater rate, and the interaction
time increases while wavepacket spreading remains negligible to order
$\sqrt{\epsilon}.$ So the small parameter for the approximations
we will use must be taken as $\sqrt{\epsilon}.$

It is puzzling why more researchers do not use wavepackets in their
calculations. Perhaps it is felt that they are an unnecessary complication.
They are certainly a complication, as we will see. However it is the
point of this paper to demonstrate that they are \textit{necessary}
if one wants to obtain finite results. A plane wave (a momentum eigenvector)
has a position probability density spread evenly over the entire universe.
This is hardly what we would call a realistic representation of a
physical situation. In the standard derivation of the cross section
\cite{Itzykson1980}, one must deal with the square of a Dirac delta
function. No such problem arises in this presentation. Probability
amplitudes can be defined in the relativistic domain \cite{Hoffmann2018d,Rosenstein1985,Fong1968,Foldy1956},
with well-defined Lorentz transformation properties.

In Section \ref{sec:Matrix-elements} we will present our results
for all terms up to second order. Conclusions follow in Section \ref{sec:Conclusions}.

Throughout this paper, we use Heaviside-Lorentz units, in which $\hbar=c=\epsilon_{0}=\mu_{0}=1$.

\section{\label{sec:Matrix-elements}Matrix elements}

The Hamiltonian we consider is, in position space,
\begin{align}
H & =H_{0}+V\nonumber \\
 & =-\frac{1}{2m_{0}}\frac{\partial^{2}}{\partial\boldsymbol{r}^{2}}+\frac{\alpha}{r},\label{eq:3}
\end{align}
with $m_{0}$ the mass, $r=|\boldsymbol{r}|$ and $\alpha$ the fine
structure constant (or $Z_{1}Z_{2}\,\alpha$ for atomic number $Z_{1}$
of the target and $Z_{2}$ of the projectile).

For nonrelativistic quantum mechanics with a potential operator, $V,$
the Born series for the full evolution operator is \cite{Messiah1961},
to second order,
\begin{equation}
e^{-iHt}=e^{-iH_{0}t}-i\int_{0}^{t}e^{-iH_{0}(t-\tau)}\,V\,e^{-iH_{0}\tau}d\tau-\int_{0}^{t}e^{-iH_{0}(t-\tau_{2})}\,V\int_{0}^{\tau_{2}}e^{-iH_{0}(\tau_{2}-\tau_{1})}\,V\,e^{-iH_{0}\tau_{1}}\,d\tau_{1}d\tau_{2}+\dots.\label{eq:4}
\end{equation}
Our initial and final momentum wavefunctions, normalized to unity,
are given by
\begin{align}
\Psi_{i}(\boldsymbol{k}) & =\frac{e^{-(\boldsymbol{k}-\boldsymbol{p}_{i})^{2}/4\sigma_{p}^{2}}}{(2\pi\sigma_{p}^{2})^{\frac{3}{4}}}e^{-i\boldsymbol{k}\cdot\boldsymbol{R}_{i}},\nonumber \\
\Psi_{f}(\boldsymbol{k}) & =\frac{e^{-(\boldsymbol{k}-\boldsymbol{p}_{f})^{2}/4\sigma_{p}^{2}}}{(2\pi\sigma_{p}^{2})^{\frac{3}{4}}}e^{-i\boldsymbol{k}\cdot\boldsymbol{R}_{f}},\label{eq:5}
\end{align}
with
\begin{align}
\boldsymbol{R}_{i} & =-R\hat{\boldsymbol{z}},\quad\boldsymbol{p}_{i}=+p\hat{\boldsymbol{z}},\nonumber \\
\boldsymbol{R}_{f} & =+R(\sin\theta,0,\cos\theta),\quad\boldsymbol{p}_{f}=+p(\sin\theta,0,\cos\theta)\label{eq:6}
\end{align}
and $\theta$ the scattering angle. The interaction time is the time
for free evolution
\begin{equation}
T=\frac{2R}{p/m_{0}}.\label{eq:7}
\end{equation}
(We do not attempt to calculate time delays or advancements here,
as was done in \cite{Hoffmann2017a}.) Note that we are representing
the special geometry of a head-on collision, as was done in \cite{Hoffmann2017a}.

We need to calculate the matrix elements
\begin{equation}
\mathcal{M}(\theta)=\langle\,\psi_{f}\,|\,e^{-iHT}\,|\,\psi_{i}\,\rangle\label{eq:8}
\end{equation}
as contributions from each order. In anticipation of our result, we
call these amplitudes \textit{finite amplitudes}, using the notation
introduced in \cite{Hoffmann2017a}. They are such that their modulus
squared is a probability. In contrast, \textit{scattering amplitudes,
$f(\theta),$} as defined in the literature \cite{Messiah1961} are
such that the differential cross section is
\begin{equation}
\frac{d\sigma}{d\Omega}=|f(\theta)|^{2}.\label{eq:9}
\end{equation}

We note that if Eq. (\ref{eq:4}) is, in fact, an approximation of
a unitary operator, then we are guaranteed to find finite contributions
by the Schwartz inequality \cite{Messiah1961}
\begin{equation}
|\langle\,\psi_{f}\,|\,e^{-iHT}\,|\,\psi_{i}\,\rangle|\leq1\label{eq:10}
\end{equation}

\subsection{Zeroth order, $\mathcal{M}_{0}(\theta)$}

First
\begin{align}
\mathcal{M}_{0} & (\theta)=\int d^{3}k_{1}\,\Psi_{f}^{*}(k_{1})\,e^{-iE_{1}T}\,\Psi_{i}(k_{1})\nonumber \\
 & =\int d^{3}k_{1}\,\frac{e^{-(\boldsymbol{k}_{1}-\boldsymbol{p}_{i})^{2}/4\sigma_{p}^{2}}}{(2\pi\sigma_{p}^{2})^{\frac{3}{4}}}\,\frac{e^{-(\boldsymbol{k}_{1}-\boldsymbol{p}_{f})^{2}/4\sigma_{p}^{2}}}{(2\pi\sigma_{p}^{2})^{\frac{3}{4}}}\,e^{+i\boldsymbol{k}_{1}\cdot(\boldsymbol{R}_{f}-\boldsymbol{R}_{i})}\,e^{-iE_{1}T},\label{eq:11}
\end{align}
with $E_{n}=\boldsymbol{k}_{n}^{2}/2m_{0}.$ We will repeatedly use
the identity
\begin{equation}
(\boldsymbol{k}-\boldsymbol{p}_{f})^{2}+(\boldsymbol{k}-\boldsymbol{p}_{i})^{2}=2(\boldsymbol{k}-\boldsymbol{p}_{+})^{2}+\frac{1}{2}(\boldsymbol{p}_{f}-\boldsymbol{p}_{i})^{2},\label{eq:12}
\end{equation}
with
\begin{equation}
\boldsymbol{p}_{+}\equiv\frac{1}{2}(\boldsymbol{p}_{f}+\boldsymbol{p}_{i}).\label{eq:13}
\end{equation}
Changing integration variables to $\boldsymbol{\kappa}=\boldsymbol{k}_{1}-\boldsymbol{p}_{+}$
involves a straightforward application of the integral \cite{Gradsteyn1980}
(their Eq. (3.323.2))
\begin{equation}
\int d^{3}\kappa\,e^{-\boldsymbol{\kappa}^{2}/2\sigma_{p}^{2}}\,e^{i\boldsymbol{\kappa}\cdot\boldsymbol{\xi}}=(2\pi\sigma_{p}^{2})^{\frac{3}{2}}\,e^{-\boldsymbol{\xi}^{2}/8\sigma_{x}^{2}},\label{eq:14}
\end{equation}
which we will use frequently. The only point to note is that the phase
factor $e^{-i\kappa^{2}T/2m_{0}}$ can be set to $1+\mathcal{O}(\sqrt{\epsilon})$
with our scaling scheme to avoid wavepacket spreading.

We find
\begin{equation}
\mathcal{M}_{0}(\theta)=e^{+iET}\,e^{-\sin^{2}\frac{\theta}{2}/2\epsilon^{2}}(1+\mathcal{O}(\sqrt{\epsilon})).\label{eq:15}
\end{equation}
Of course this is a function sharply peaked around zero scattering
angle. Note that the modulus-squared of the amplitude $\mathcal{M}$
we calculate is a probability, and this probability is 1 for $\theta=0.$

\subsection{First order, $\mathcal{M}_{1}(\theta)$}

We want to evaluate the matrix elements in momentum space, as the
free evolution operators take their simplest representation, as phase
factors, in this basis. Multiplication of the position wavefunction
by the factor $\alpha/r$ from the Coulomb potential is equivalent
to convolution of the momentum wavefunction. We find
\begin{align}
\Psi^{\prime}(\boldsymbol{k}) & =\langle\,\boldsymbol{k}\,|\,\frac{\alpha}{r}\,|\,\psi\,\rangle=\int\frac{d^{3}r}{(2\pi)^{\frac{3}{2}}}e^{-i\boldsymbol{k}\cdot\boldsymbol{r}}\,\frac{\alpha}{r}\,\int\frac{d^{3}k^{\prime}}{(2\pi)^{\frac{3}{2}}}e^{i\boldsymbol{k}^{\prime}\cdot\boldsymbol{r}}\,\Psi(\boldsymbol{k}^{\prime}).\label{eq:16}
\end{align}
Exchanging the order of the integrals gives
\begin{equation}
\Psi^{\prime}(\boldsymbol{k})=\int d^{3}k^{\prime}\,\mathcal{V}(\boldsymbol{k}-\boldsymbol{k}^{\prime})\,\Psi(\boldsymbol{k}^{\prime}),\label{eq:17}
\end{equation}
with
\begin{align}
\mathcal{V}(\boldsymbol{q}) & =\int\frac{d^{3}r}{(2\pi)^{3}}\,e^{-i\boldsymbol{q}\cdot\boldsymbol{r}}\,\frac{\alpha}{r}=\frac{\alpha}{2\pi^{2}q}\int_{0}^{\infty}dr\,\sin(qr).\label{eq:18}
\end{align}
Integration in spherical polar coordinates cancels the zero of the
denominator, a technique that we will use many times. Note that the
integral for $\mathcal{V},$ as written, diverges for all $q.$ The
exchange of the order of integration was not justified. We interpret
$\mathcal{V}$ as a distribution \cite{Richard1995}. If it is made
finite by a regularization procedure, it can be used in integrals.
Then, finally, the regularization is removed to give, possibly, a
finite result. This procedure is what we are doing when we use Dirac
delta functions. In this case, it is equivalent to what we would find
if we kept the original integration order.

We insert a Yukawa regularization factor, $e^{-\lambda r},$ in the
last form of Eq. (\ref{eq:18}) to find the regularized $\bar{\mathcal{V}}$
\begin{equation}
\bar{\mathcal{V}}(\boldsymbol{q})=\frac{\alpha}{2\pi^{2}}\,\frac{1}{\boldsymbol{q}^{2}+\lambda^{2}}.\label{eq:19}
\end{equation}
We will refer to this as the potential kernel.

Then the first order matrix element is
\begin{equation}
\mathcal{M}_{1}(\theta)=-i\int_{0}^{T}d\tau\int d^{3}k\,\Psi_{f}^{*}(\boldsymbol{k})\,e^{-iE(\boldsymbol{k})(T-\tau)}\,\frac{\alpha}{2\pi^{2}}\int d^{3}k^{\prime}\,\frac{1}{(\boldsymbol{k}-\boldsymbol{k}^{\prime})^{2}+\lambda^{2}}\,e^{-iE(\boldsymbol{k}^{\prime})\tau}\,\Psi_{i}(\boldsymbol{k}^{\prime}).\label{eq:20}
\end{equation}
We can evaluate this quite simply for scattering sufficiently far
from the forward direction. In that region, the Gaussians keep $\boldsymbol{k}$
close to $\boldsymbol{p}_{f}$ and $\boldsymbol{k}^{\prime}$ close
to $\boldsymbol{p}_{i}.$ The potential kernel is slowly varying,
so can be replaced by its value at the wavefunction peaks, with no
singularity, and taken outside the integral. The remaining integrals
are easily evaluated using the methods of the previous section.

We find
\begin{equation}
\mathcal{M}_{1}(\theta)=-i\,e^{+iET}\,\alpha\,\frac{\sigma_{p}^{2}}{p}\,\frac{1}{E\sin^{2}\frac{\theta}{2}}=-i\,e^{+iET}\,2\eta\,\epsilon^{2}\,\frac{1}{\sin^{2}\frac{\theta}{2}},\quad\mathrm{for}\ \theta\neq0,\label{eq:21}
\end{equation}
where
\begin{equation}
\eta=\frac{\alpha}{p/m_{0}}\label{eq:22}
\end{equation}
is a commonly used dimensionless measure of the strength of the Coulomb
interaction \cite{Messiah1961}.

Then we use a formula from \cite{Hoffmann2017a}, derived for Gaussian
wavepackets, that relates probabilities, $P,$ and differential cross
sections:
\begin{equation}
\frac{d\sigma}{d\Omega}=\frac{p^{2}}{16\sigma_{p}^{4}}\,P.\label{eq:23}
\end{equation}
Since $\mathcal{M}_{0}(\theta)$ essentially vanishes in this region,
we have
\begin{align}
\frac{d\sigma}{d\Omega} & =\frac{p^{2}}{16\sigma_{p}^{4}}\,|\mathcal{M}_{1}|^{2}=\frac{\alpha^{2}}{16\,E^{2}\sin^{4}\frac{\theta}{2}}.\label{eq:24}
\end{align}

This is the well-known Rutherford cross section \cite{Rutherford1911},
known to be in good agreement with experiment in the nonrelativistic
regime and where spin contributions are negligible \cite{Geiger1909}.
It is possible that the dependence on the wavepacket shape cancels
out to give this expression. Investigation of that possibility must
await a future work.

\subsection{First order in the forward direction, $\mathcal{M}_{1}(0)$}

For forward scattering, in Eq. (\ref{eq:20}), we set $\boldsymbol{p}_{f}=\boldsymbol{p}_{i}=p\,\hat{\boldsymbol{z}}$
and $\boldsymbol{R}_{f}=-\boldsymbol{R}_{i}=R\,\hat{\boldsymbol{z}}.$
While the integrals are regularized with a nonzero value of $\lambda,$
we can change integration variables to $\boldsymbol{k}_{-}=\boldsymbol{k}-\boldsymbol{k}^{\prime}$
and $\boldsymbol{k}_{+}=\boldsymbol{k}+\boldsymbol{k}^{\prime},$
with $d^{3}k\,d^{3}k^{\prime}=\frac{1}{8}d^{3}k_{+}d^{3}k_{-}.$ In
the integral over $\boldsymbol{k}_{-},$ $\lambda$ can now be set
to zero as the integration measure in spherical polar coordinates
cancels the divergence. We find, using M{\footnotesize{}ATHEMATICA
}\cite{Mathematica2019},
\begin{equation}
\int d^{3}k_{-}\,e^{-k_{-}^{2}/8\sigma_{p}^{2}}\,\frac{1}{k_{-}^{2}+\lambda^{2}}\,e^{i2\boldsymbol{v}_{+}\cdot\boldsymbol{k}_{-}(\tau-\frac{T}{2})}\rightarrow\pi^{2}\,\frac{\mathrm{erf}\,(\sqrt{2}\,v_{+}(\tau-\frac{T}{2})/\sigma_{x})}{\,v_{+}(\tau-\frac{T}{2})},\label{eq:24.1}
\end{equation}
with $v_{+}=k_{+}/m_{0}.$ The argument of this function varies by
only $\mathcal{O}(\sqrt{\epsilon})$ as $k_{+}$ changes by $\sigma_{p},$
while it changes by order unity as $\tau$ changes by order $\sqrt{\epsilon}.$
The function is peaked at the origin with a width of order unity.
So, since the function is slowly varying with respect to momentum,
it can be set to its value at $k_{+}=2\,\boldsymbol{p}$ and taken
outside the $k_{+}$ integral.

This gives
\begin{equation}
\mathcal{M}_{1}(0)=-i\frac{\alpha}{2\pi^{2}}\frac{1}{(2\pi\sigma_{p}^{2})^{\frac{3}{2}}}\frac{1}{8}\int_{0}^{T}d\tau\,\pi^{2}\,\frac{\mathrm{erf}\,(\sqrt{2}\,2v(\tau-\frac{T}{2})/\sigma_{x})}{2v(\tau-\frac{T}{2})}\int d^{3}k_{+}\,e^{+i\boldsymbol{k}_{+}\cdot\boldsymbol{R}_{f}}\,e^{-(\boldsymbol{k}_{+}-2\boldsymbol{p})^{2}/8\sigma_{p}^{2}}\,e^{-iE_{+}T/4}.\label{eq:24.2}
\end{equation}
Evaluating the $k_{+}$ integral gives
\begin{equation}
\mathcal{M}_{1}(0)=-e^{+iET}\,i\frac{\eta}{4}\,\int_{-Z}^{Z}dz\,\frac{\mathrm{erf}\,(z)}{z},\label{eq:24.3}
\end{equation}
with $Z=\sqrt{8/\epsilon}.$ Numerical integration with M{\footnotesize{}ATHEMATICA
}\cite{Mathematica2019} gives an approximate expression
\begin{equation}
\mathcal{M}_{1}(0)\cong-e^{+iET}\,i\,\frac{\eta}{4}\,(2.00\,\ln(\sqrt{\frac{8}{\epsilon}})+1.96).\label{eq:24.4}
\end{equation}

We see that this expression contains the same phase factor, $\exp(+iET),$
as $\mathcal{M}_{0}(0).$ It is proportional to $\eta$ with a factor
that depends only logarithmically on $\epsilon.$ For example, with
$\epsilon=0.001,$ it takes the form
\begin{equation}
\mathcal{M}_{1}(0)\cong-e^{+iET}\,i\,2.74\,\eta.\label{eq:24.5}
\end{equation}

\subsection{Second order, $\mathcal{M}_{2}(\theta)$}

The second order matrix element is
\begin{multline}
\mathcal{M}_{2}(\theta)=-\int_{0}^{T}d\tau_{2}\int_{0}^{\tau_{2}}d\tau_{1}\int d^{3}k_{3}\,\Psi_{f}^{*}(\boldsymbol{k}_{3})\,e^{-iE_{3}(T-\tau_{2})}\int d^{3}k_{2}\,\frac{\alpha}{2\pi^{2}}\,\frac{1}{(\boldsymbol{k}_{3}-\boldsymbol{k}_{2})^{2}+\lambda^{2}}\,e^{-iE_{2}(\tau_{2}-\tau_{1})}\times\\
\times\int d^{3}k_{1}\,\frac{\alpha}{2\pi^{2}}\,\frac{1}{(\boldsymbol{k}_{2}-\boldsymbol{k}_{1})^{2}+\lambda^{2}}\,e^{-iE_{1}\tau_{1}}\,\Psi_{i}(\boldsymbol{k}_{1}).\label{eq:26}
\end{multline}

If we tried to evaluate the intermediate $\boldsymbol{k}_{2}$ integral
first, with two potential kernel factors
\begin{equation}
D=\frac{1}{(\boldsymbol{k}_{2}-\boldsymbol{k}_{3})^{2}}\frac{1}{(\boldsymbol{k}_{2}-\boldsymbol{k}_{1})^{2}}\label{eq:27}
\end{equation}
in the integrand, we would not be able to cancel divergences at both
$\boldsymbol{k}_{3}$ and $\boldsymbol{k}_{1}.$ So we perform the
$\boldsymbol{k}_{1}$ integral first, then the $\boldsymbol{k}_{3}$
integral and finally the $\boldsymbol{k}_{2}$ integral.

First we evaluate
\begin{equation}
I_{1}=\int d^{3}k_{1}\,\frac{1}{(\boldsymbol{k}_{1}-\boldsymbol{k}_{2})^{2}+\lambda^{2}}\,e^{-iE_{1}\tau_{1}}\,\Psi_{i}(\boldsymbol{k}_{1}).\label{eq:28}
\end{equation}
While the potential kernel is regularized, we can change integration
variables to $\boldsymbol{\kappa}=\boldsymbol{k}_{1}-\boldsymbol{k}_{2}.$
Then we can set $\lambda=0$ since, in spherical polar coordinates,
the factor $\kappa^{2}$ from the integration measure cancels the
factor $\kappa^{2}$ in the denominator. We then find
\begin{equation}
I_{1}=\frac{1}{(2\pi\sigma_{p}^{2})^{\frac{3}{4}}}e^{-iE_{2}\tau_{1}}e^{-i\boldsymbol{k}_{2}\cdot\boldsymbol{R}_{i}}e^{-(\boldsymbol{k}_{2}-\boldsymbol{p}_{i})^{2}/4\sigma_{p}^{2}}\int_{0}^{\infty}d\kappa\int d^{2}\hat{\boldsymbol{\kappa}}\,e^{-\kappa^{2}/4\sigma_{p}^{2}}\,e^{-iE(\boldsymbol{\kappa})\tau_{1}}\,e^{\boldsymbol{\kappa}\cdot\boldsymbol{\xi}_{i}},\label{eq:29}
\end{equation}
where
\begin{equation}
\boldsymbol{\xi}_{i}=-\frac{\boldsymbol{k}_{2}-\boldsymbol{p}_{i}}{2\sigma_{p}^{2}}-i(\boldsymbol{R}_{i}+\boldsymbol{v}(\boldsymbol{k}_{2})\tau_{1}).\label{eq:30}
\end{equation}
We will use the notation
\begin{equation}
E(\boldsymbol{k})=\frac{\boldsymbol{k}^{2}}{2m_{0}}\quad\mathrm{and}\quad\boldsymbol{v}(\boldsymbol{k})=\frac{\boldsymbol{k}}{m_{0}}.\label{eq:31}
\end{equation}

The factor $\exp(-(\boldsymbol{k}_{2}-\boldsymbol{p}_{i})^{2}/4\sigma_{p}^{2})$
will not, in fact, survive once we find the dependence on $\boldsymbol{k}_{2}$
of the remaining integrals. But another factor will emerge that regularizes
the $\boldsymbol{k}_{2}$ integral, as we will see below. Remarkably,
the regularizing influence of the initial wavefunction is extended
to the intermediate $\boldsymbol{k}_{2}$ integral.

We know that
\begin{equation}
\int d^{2}\hat{\boldsymbol{\kappa}}\,e^{\boldsymbol{A}\cdot\boldsymbol{\kappa}}=4\pi\,\frac{\sinh(\sqrt{\boldsymbol{A}^{2}}\,\kappa)}{\sqrt{\boldsymbol{A}^{2}}\,\kappa}\quad\mathrm{and}\quad\int d^{2}\hat{\boldsymbol{\kappa}}\,e^{iB\cdot\kappa}=4\pi\,\frac{\sin(\sqrt{\boldsymbol{B}^{2}}\,\kappa)}{\sqrt{\boldsymbol{B}^{2}}\,\kappa}.\label{eq:32}
\end{equation}
These functions are both even and entire, with power series that contain
only integral powers of $\boldsymbol{A}^{2}$ and $\boldsymbol{B}^{2},$
respectively. If we put $\boldsymbol{A}=i\boldsymbol{B}$ in the first
formula, we get the second formula. So, by analytic continuation,
\begin{equation}
\int d^{2}\hat{\boldsymbol{\kappa}}\,e^{(\boldsymbol{A}+i\boldsymbol{B})\cdot\boldsymbol{\kappa}}=4\pi\,\frac{\sinh(\sqrt{(\boldsymbol{A}+i\boldsymbol{B})^{2}}\,\kappa)}{\sqrt{(\boldsymbol{A}+i\boldsymbol{B})^{2}}\,\kappa}.\label{eq:33}
\end{equation}

In the remaining integral, we can replace
\begin{equation}
e^{-iE(\boldsymbol{\kappa})\tau_{1}}\rightarrow1+\mathcal{O}(\sqrt{\epsilon})\label{eq:34}
\end{equation}
within the peak region of the Gaussian $\exp(-\kappa^{2}/4\sigma_{p}^{2}).$
Then we have
\begin{equation}
\int_{0}^{\infty}d\kappa\,e^{-\kappa^{2}/4\sigma_{p}^{2}}\,\frac{\sinh(\sqrt{\boldsymbol{\xi}_{i}{}^{2}}\,\kappa)}{\sqrt{\boldsymbol{\xi}_{i}^{2}}\,\kappa}=\frac{1}{\sigma_{x}}\frac{\sqrt{\pi}}{2}\,g(\frac{\boldsymbol{\xi}_{i}^{2}}{4\sigma_{x}^{2}}),\label{eq:35}
\end{equation}
with
\begin{equation}
g(x)\equiv\frac{2}{\sqrt{\pi}}\int_{0}^{\infty}dz\,e^{-z^{2}}\,\frac{\sinh(\sqrt{4x}\,z)}{\sqrt{4x}\,z}=\frac{\sqrt{\pi}}{2}\,\frac{\mathrm{erf}\,(i\sqrt{x})}{i\sqrt{x}},\label{eq:36}
\end{equation}
normalized to $g(0)=1.$ The last integral was evaluated using M{\footnotesize{}ATHEMATICA
}\cite{Mathematica2019}.

So our result for $I_{1}$ is
\begin{equation}
I_{1}=\frac{4\pi}{(2\pi\sigma_{p}^{2})^{\frac{3}{4}}}e^{-iE_{2}\tau_{1}}e^{-i\boldsymbol{k}_{2}\cdot\boldsymbol{R}_{i}}e^{-(\boldsymbol{k}_{2}-\boldsymbol{p}_{i})^{2}/4\sigma_{p}^{2}}\,\frac{1}{\sigma_{x}}\frac{\sqrt{\pi}}{2}\,g(\frac{\boldsymbol{\xi}_{i}^{2}}{4\sigma_{x}^{2}}).\label{eq:37}
\end{equation}

Similarly, we find
\begin{align}
I_{3} & =\int d^{3}k_{3}\,\frac{1}{(\boldsymbol{k}_{3}-\boldsymbol{k}_{2})^{2}}\,e^{-iE_{3}(T-\tau_{2})}\,\Psi_{f}^{*}(\boldsymbol{k}_{3})\nonumber \\
 & =\frac{4\pi}{(2\pi\sigma_{p}^{2})^{\frac{3}{4}}}e^{-iE_{2}(T-\tau_{2})}e^{+i\boldsymbol{k}_{2}\cdot\boldsymbol{R}_{f}}e^{-(\boldsymbol{k}_{2}-\boldsymbol{p}_{f})^{2}/4\sigma_{p}^{2}}\,\frac{1}{\sigma_{x}}\frac{\sqrt{\pi}}{2}\,g^{*}(\frac{\boldsymbol{\xi}_{f}^{2}}{4\sigma_{x}^{2}}),\label{eq:38}
\end{align}
with
\begin{equation}
\boldsymbol{\xi}_{f}=-\frac{\boldsymbol{k}_{2}-\boldsymbol{p}_{f}}{2\sigma_{p}^{2}}-i(\boldsymbol{R}_{f}+\boldsymbol{v}(\boldsymbol{k}_{2})(T-\tau_{2})).\label{eq:39}
\end{equation}

Then we find
\begin{multline}
\mathcal{M}_{2}(\theta)=-\frac{\alpha^{2}}{\pi\sigma_{x}^{2}}\int_{0}^{T}d\tau_{2}\int_{0}^{\tau_{2}}d\tau_{1}\int d^{3}k_{2}\,\frac{1}{(2\pi\sigma_{p}^{2})^{\frac{3}{2}}}\,e^{-(\boldsymbol{k}_{2}-\boldsymbol{p}_{f})^{2}/4\sigma_{p}^{2}}e^{-(\boldsymbol{k}_{2}-\boldsymbol{p}_{i})^{2}/4\sigma_{p}^{2}}\times\\
\times e^{-iE_{2}T}\,e^{i\boldsymbol{k}_{2}\cdot(\boldsymbol{R}_{f}-\boldsymbol{R}_{i})}\,g^{*}(\frac{\boldsymbol{\xi}_{f}^{2}}{4\sigma_{x}^{2}})g(\frac{\boldsymbol{\xi}_{i}^{2}}{4\sigma_{x}^{2}}).\label{eq:40}
\end{multline}
Combining Gaussian exponents using Eq. (\ref{eq:12}) gives
\begin{equation}
\mathcal{M}_{2}(\theta)=-\frac{\alpha^{2}}{\pi\sigma_{x}^{2}}\,e^{-(\boldsymbol{p}_{f}-\boldsymbol{p}_{i})^{2}/8\sigma_{p}^{2}}\,\int_{0}^{T}d\tau_{2}\int_{0}^{\tau_{2}}d\tau_{1}\int d^{3}k_{2}\,\frac{e^{-(\boldsymbol{k}_{2}-\boldsymbol{p}_{+})^{2}/2\sigma_{p}^{2}}}{(2\pi\sigma_{p}^{2})^{\frac{3}{2}}}\,e^{-iE_{2}T}\,e^{i\boldsymbol{k}_{2}\cdot(\boldsymbol{R}_{f}-\boldsymbol{R}_{i})}\,g^{*}(\frac{\boldsymbol{\xi}_{f}^{2}}{4\sigma_{x}^{2}})g(\frac{\boldsymbol{\xi}_{i}^{2}}{4\sigma_{x}^{2}}),\label{eq:41}
\end{equation}
with $\boldsymbol{p}_{+}$ as in Eq. (\ref{eq:13}). Then we change
variables to $\boldsymbol{\kappa}=\boldsymbol{k}_{2}-\boldsymbol{p}_{+}.$

We use two properties of the $g$ function that we have verified numerically.
The arguments of the $g$ function, $\boldsymbol{\xi}_{i}^{2}/4\sigma_{x}^{2}$
and $\boldsymbol{\xi}_{f}^{2}/4\sigma_{x}^{2},$ both contain a large,
real, positive term ($M=\sin^{2}\frac{\theta}{2}/4\epsilon^{2}$ for
$|\theta|\gg\epsilon$ in our case) and other terms that remain of
order unity. We find
\begin{equation}
g(M)\rightarrow\frac{e^{M}}{\pi M}\quad\mathrm{for}\ M\gg1.\label{eq:42}
\end{equation}
Then for proportionally small fluctuations around that large, positive
value,
\begin{equation}
\frac{g(M+z)}{g(M)}\rightarrow e^{z}\quad\mathrm{for}\ M\gg1\ \mathrm{and}\ |z|\ll M.\label{eq:43}
\end{equation}
These two results are consistent.

Using these approximations gives
\begin{multline}
g^{*}(\frac{\boldsymbol{\xi}_{f}^{2}}{4\sigma_{x}^{2}})g(\frac{\boldsymbol{\xi}_{i}^{2}}{4\sigma_{x}^{2}})\rightarrow\frac{4\epsilon^{4}e^{\sin^{2}\frac{\theta}{2}/2\epsilon^{2}}}{\pi^{4}\sin^{4}\frac{\theta}{2}}e^{+\kappa^{2}/2\sigma_{p}^{2}}e^{-\sin^{2}\frac{\theta}{2}/2\epsilon}e^{-\cos^{2}\frac{\theta}{2}(\tau_{1}+\tau_{2}-T){}^{2}/2\epsilon T^{2}}\times\\
\times e^{-(\boldsymbol{\kappa}+\boldsymbol{p}_{+}(\tau_{1}-\tau_{2})/T)^{2}/2\epsilon p^{2}}e^{i\sin^{2}\frac{\theta}{2}/\epsilon^{\frac{3}{2}}}e^{i\boldsymbol{\kappa}\cdot\boldsymbol{\chi}},\label{eq:44}
\end{multline}
with
\begin{equation}
\boldsymbol{\chi}=\boldsymbol{v}_{f}(\tau_{1}-\frac{T}{2})-\boldsymbol{v}_{i}(\tau_{2}-\frac{T}{2}).\label{eq:45}
\end{equation}
We note the cancellations
\begin{equation}
e^{-(\boldsymbol{p}_{f}-\boldsymbol{p}_{i})^{2}/8\sigma_{p}^{2}}\,e^{+\sin^{2}\frac{\theta}{2}/2\epsilon^{2}}=1\quad\mathrm{and}\quad e^{-(\boldsymbol{k}_{2}-\boldsymbol{p}_{+})^{2}/2\sigma_{p}^{2}}\,e^{+\kappa^{2}/2\sigma_{p}^{2}}=1.\label{eq:46}
\end{equation}
Changing variables again, to $\boldsymbol{\kappa}^{\prime}=\boldsymbol{\kappa}+\boldsymbol{p}_{+}(\tau_{1}-\tau_{2})/T,$
gives a remaining constraining factor $\exp(-\kappa^{\prime2}/2\epsilon p^{2}),$
as mentioned earlier. The width of this function is $\sqrt{\epsilon}\,p,$
wider than the previous $\epsilon\,p=\sigma_{p}.$

The remaining $\boldsymbol{\kappa}^{\prime}$ integral is performed
using Eq. (\ref{eq:14}). Then, for the time integrals, we use a change
of variables and extend some of the limits, since the widths of the
distributions centred on $\tau_{1}=T/2$ and $\tau_{2}=T/2$ are of
order $\epsilon T,$ much smaller than the widths of the integration
regions.
\begin{equation}
\int_{0}^{T}d\tau\int_{0}^{\tau_{2}}d\tau_{1}\rightarrow\frac{1}{2}\,T^{2}\,\int_{-\infty}^{\infty}dx_{+}\int_{-\infty}^{0}dx_{-},\label{eq:47}
\end{equation}
with
\begin{equation}
x_{\pm}=\frac{\tau_{1}-\frac{T}{2}}{T}\pm\frac{\tau_{2}-\frac{T}{2}}{T}.\label{eq:48}
\end{equation}
Finally, using Eq. (\ref{eq:14}) twice in the case $\boldsymbol{\xi}=0$
(the $x_{-}$ integral is the integral of an even function on the
half line, equal to one half the integral on the full line),
\begin{equation}
\int_{-\infty}^{\infty}dx_{+}\int_{-\infty}^{0}dx_{-}\,e^{-\sin^{2}\frac{\theta}{2}x_{+}^{2}/2\epsilon^{2}}\,e^{-2\cos^{2}\frac{\theta}{2}x_{-}^{2}/\epsilon^{2}}=(\frac{2\pi\epsilon^{2}}{\sin^{2}\frac{\theta}{2}})^{\frac{1}{2}}\frac{1}{2}(\frac{\pi\epsilon^{2}}{2\cos^{2}\frac{\theta}{2}})^{\frac{1}{2}}.\label{eq:49}
\end{equation}

Our final result is
\begin{equation}
\mathcal{M}_{2}(\theta)=-e^{+iET}\,e^{i\sin^{2}\frac{\theta}{2}/2\epsilon^{\frac{3}{2}}}\,\frac{16}{\pi^{2}}\eta^{2}\epsilon^{\frac{7}{2}}\frac{e^{-\sin^{2}\frac{\theta}{2}/2\epsilon}}{\sin^{5}\frac{\theta}{2}\cos\frac{\theta}{2}}\quad\mathrm{for}\ \theta\neq0.\label{eq:50}
\end{equation}
The presence of a half-integral power of $\epsilon$ comes about because
the width of the distribution $\exp(-\kappa^{\prime2}/2\epsilon p^{2})$
is $\sqrt{\epsilon}\,p.$

We see that this term gives a correction to $\mathcal{M}_{1}(\theta)$
that is negligible everywhere except close to the forward direction,
where the approximations used to derive this term become no longer
valid.

\subsection{Second order in the forward direction, $\mathcal{M}_{2}(0)$}

We fully expect that calculation of the second order contribution
in the forward direction will lead to a finite result. The calculation
is challenging and will be left for future work. From what we saw
for $\mathcal{M}_{1}(0),$ we expect the result to be proportional
to the same phase factor $\exp(+iET),$ so that it can interfere with
$\mathcal{M}_{0}(0),$ and to be proportional to $\eta^{2}$ with
a factor that is neither very large nor very small and depends only
weakly on $\epsilon.$ In the calculation of $\mathcal{M}_{1}(0),$
we accomplished our intention to show that this forward amplitude
is finite and has the correct dependence on the parameters.

In \cite{Hoffmann2017a}, we used a wavepacket treatment that made
the partial wave series converge for Coulomb scattering. We found
that for $\eta=\pm1,$ the amplitude in the forward direction was
very small. It will be of value to calculate $\mathcal{M}_{2}(0)$
using the methods of this paper to allow estimation of the probability
for forward scattering to order $\eta^{2}.$ This will allow comparison
between these two very different calculation methods for the same
physical description, and to potentially observe a falloff of the
probability of forward scattering with increasing $\eta.$

Probabilities in the forward direction are not experimentally measurable.
As discussed in \cite{Hoffmann2017a}, in an experiment such as a
beam of alpha particles incident on a gold foil, projectiles with
sufficiently large impact parameters relative to the nuclei are expected
to pass by largely undisturbed and contribute to a strong peak of
flux around the forward direction. This would obscure any attempt
to measure probability in the forward direction. The calculation presented
in this paper used the particular geometry of zero impact parameter.

\section{\label{sec:Conclusions}Conclusions}

If the matrix element for the second-order Born approximation to nonrelativistic
Coulomb scattering is calculated between plane-wave state vectors
(momentum eigenvectors), the result diverges. Calculating the matrix
element between square-integrable wavepacket state vectors gives results
that are everywhere finite, including in the forward direction, and
with the correct physical properties. We found that the finite second-order
term makes a negligible contribution to the scattering away from the
forward direction. This is reassuring, as the first-order result already
gives an excellent agreement with experiment and the exact result
for the cross section is just the Rutherford formula \cite{Messiah1961}.

If these results can be extended to relativistic quantum field theory,
quantum electrodynamics for example, it will greatly change the character
of the renormalization program. The physical essence of renormalization
is that the mass and charge input to the theory acquire corrections
to become the observed, physical, mass and charge. The results presented
here suggest that this might be done with finite corrections, rather
than by manipulating regularized infinities.

Whether this is the case must await future investigation.

\bibliographystyle{vancouver}

\end{document}